# Underlying burning resistant mechanisms for titanium alloy

Yongnan Chen[a, b,*], Wenqing Yang [a]，Arixin Bo[b], Haifei Zhan[b,c,*], Fengying Zhang[a], Yongqing Zhao[d], Qinyang Zhao[e], Mingpan Wan[f], and Yuantong Gu[b]

[a] School of materials science and engineering, Chang'an University, Xi'an, 710064, China

[b] School of Chemistry, Physics and Mechanical Engineering, Queensland University of Technology (QUT), Brisbane QLD 4001, Australia.

[c] School of Computing, Engineering and Mathematics, Western Sydney University, Locked Bag 1797, Penrith NSW 2751, Australia

[d] Northwest Institute for nonferrous metal research, Xi'an, 710016, China

[e] Waikato Centre for Advanced Materials, School of Engineering, The University of Waikato, Hamilton 3204, New Zealand

[f] College of Materials and Metallurgy, Guizhou University, Guiyang, 550025, China

[*]**Corresponding authors at**: School of materials science and engineering, Chang'an University, Xi'an 710064, China (Y. Chen); School of Chemistry, Physics and Mechanical Engineering, Queensland University of Technology (QUT), Brisbane, QLD 4001, Australia (H. Zhan).

E-mail addresses: frank_cyn@163.com (Y. Chen); zhan.haifei@qut.edu.au (H. Zhan)

**ABSTRACT**

The "titanium fire" as produced during high pressure and friction is the major failure scenario for aero-engines. To alleviate this issue, Ti-V-Cr and Ti-Cu-Al series burn resistant titanium alloys have been developed. However, which burn resistant alloy exhibit better property with reasonable cost needs to be evaluated. This work unveils the burning mechanisms of these alloys and discusses whether burn resistance of Cr and V can be replaced by Cu, on which thorough exploration is lacking. Two representative burn resistant alloys are considered, including Ti14 (Ti-13Cu-1Al-0.2Si) and Ti40(Ti-25V-15Cr-0.2Si) alloys. Compared with the commercial non-burn resistant titanium alloy, i.e., TC4 (Ti-6Al-4V) alloy, it has been found that both Ti14 and Ti40 alloys form "protective" shields during the burning process. Specifically, for Ti14 alloy, a clear Cu-rich layer is formed at the interface between burning product zone and heat affected zone, which consumes oxygen by producing Cu-O compounds and impedes the reaction with Ti-matrix. This work has established a fundamental understanding of burning resistant mechanisms for titanium alloys. Importantly, it is found that Cu could endow titanium alloys with similar burn resistant capability as that of V or Cr, which opens a cost-effective avenue to design burn resistant titanium alloys.

**KEYWORDS:** Titanium alloys; Burn resistance; Burning mechanism; Diffusion; Oxidation

# 1 Introduction

Titanium and its alloys are broadly applied in advanced aero-engines owing to their high thrust-weight ratio and excellent corrosion resistance. However, the so-called "titanium fire" occurs under high pressure and friction ignites unfavorable and rapid burning, which is hard to control and leads to catastrophic accidents [1-3]. To address this critical issue, tremendous attempts have been made to reduce friction, or develop burn resistant coating and burn resistant alloys in advanced aero-engine, gas industry and automobile transportation [4-6]. Since early 1970s, researchers in America and Russia had evaluated burning characteristics of titanium alloys such as Ti3515, Ti64 and Alloy C by using laser ignition, and revealed the functional relations between combustion products and gas flow conditions. In last decade, there are some researches on burn resistant property of Ti40 alloy by friction ignition method，and established the relationship between ignition temperature and pressure of Ti40 alloy.

Although several studies already reported the burn resistant performance of Ti40 alloy [7-10], wide engineering applications are very limited due to its low formability, instability at high temperature, high cost, with its burn resistant mechanisms still under study[11-13]. Recent years, researchers have turned to alternative Ti-Cu-Al series burn resistant titanium alloys, such as BTT-1(Ti-13Cu-4Al-4Mo-2Zr), BTT-3(Ti-18Cu-2Al-2Mo)[2,3] and Ti14(Ti-13Cu-1Al-0.2Si) [9]. Specifically, Ti14 alloy belongs to a new $α+Ti_2Cu$ type burn resistant titanium alloy, which have the burn resistant functionality [14] and comprise the low-cost alloying element Cu. Through the direct current simulation burning (DCSB) method [15] and metal droplet method (MDM) [16], researchers have studied the burning characteristics of Ti-Cu-Al series burn resistant titanium alloys during stable burning stage, e.g., the burning velocity and duration. However, the effect of Cu element on impeding the burning of the alloy needs to be further explained, due to the fact that the existing test methods cannot fully represent burning characteristics and more parameters needed to be considered while investigating the burn resistant mechanism of the alloys in various gas conditions.

Further investigation is needed on whether the burn resistivity provided by Cu are the same as V and Cr. Alternatively, whether Cu can totally replace V or Cr to achieve cost-effective burn resistant titanium alloy. This work aims to address these questions by studying the burning behaviors of two representative alloys through a serial of modified direct current simulation burning tests, including Ti14 and Ti40 alloys. The burning characteristics, e.g., flame height, burning duration, burning velocity, burned sample structure is performed to discuss the effect of element composition during burning process and derive the burn resistant property of the alloys. This can be an effective way to achieve competitive burn resistant function while using titanium alloys with lower cost.

## 2 Experimental Methods

### 2.1 Materials

Two typical burn resistant titanium alloys used in this study are Ti14 [17] and Ti40 [18] alloys. Raw materials of Ti40 ( Ti, 25wt%V, 15wt%Cr and 0.2wt%Si ), Ti14 (Ti, 13wt%Cu, 1wt%Al and 0.2wt%Si) are mixed to form electrode, respectively, and 25kg ingots of Ti40 and Ti14 alloy are fabricated by triple consumable vacuum arc melting, Ti40 and Ti14 alloys are conventional forged at 950℃ followed by water cooling and air cooling into a bar shape with a diameter of 40 mm to be tested, respectively [9]. The forging rate is 500mm/min and heating rate is 25℃/s. In order to determine the difference in burning behavior and burn resistant mechanism caused by V, Cr and Cu, commercial TC4 alloy (in Fig. S1) were also tested for comparison under the same testing standard. Microstructural characterization of the alloys before burning was analyzed by optical microscopy (Axio Scope-A1 equipped with a Carl Zeiss digital camera). The forged Ti14 alloy mainly consisted of α-Ti matrix and $Ti_2Cu$ precipitates (Fig. 1a) which are indicated by the white arrow (according to Ti-Cu phase diagram in Fig S2) [19]. The formation of $Ti_2Cu$ in Cu rich Ti alloy matrix is in consistent with literature [16,17]. In comparison, Ti40 alloy exhibit *β*-Ti matrix microstructure where large grains with clear grain boundaries can be observed (Fig. 1b). Average grain size of Ti14 and Ti40 alloy are 800-900μm and 200-300μm, respectively. In order to explain microstructure during burning process, phase transformation was investigated by thermodilatometry (DIL-805A/D, Bähr, Germany) and controlling temperature precision is kept as ±0.1℃. Since melting point of titanium alloy is low, samples were heated to 1100℃. Samples of Ti14 alloy experienced peritectoid transformation to form *β* phase at 780℃ and eutectic transformation to transform into liquid+*β* between 830℃ and 1070℃, while no phase change is observed in Ti40 alloy.

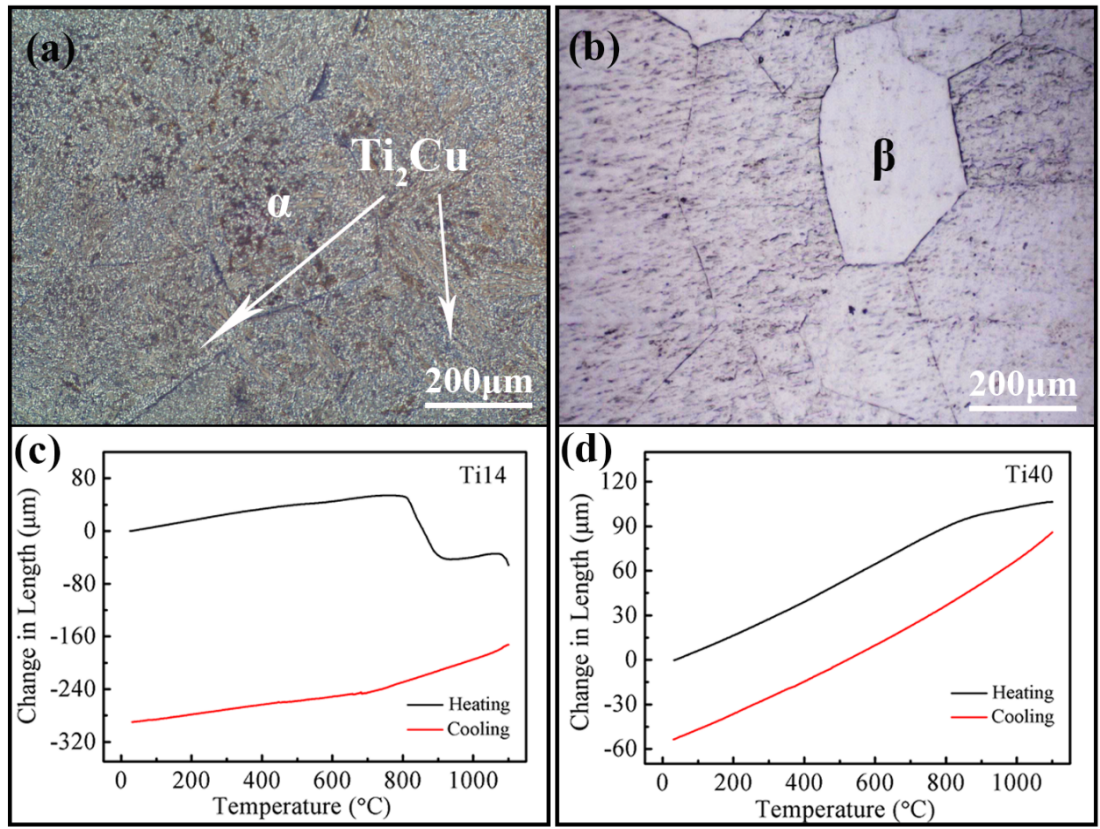

**Fig. 1**. Microstructure of Ti alloy. (a) Ti14 alloy. The $Ti_2Cu$ phase in Ti14 alloy is highlighted by white arrows; (b) Ti40 alloy. Clear grain boundaries in Ti40 alloy are shown. Phase change during heating and cooling process for: (c) Ti14 alloy, and (d) Ti40 alloy.

*2.2 Controlled burning Test*

To represent burning characteristics with various gas condition, a modified direct current simulation burning (DCSB) method was employed to study the burning behavior of the alloys [16,17]. Alloys were ignited by placing under a certain voltage and current. Meanwhile, a mixed gas ($O_2/N_2$) controller was applied to study the effect of oxygen content on burning behavior. The oxygen and nitrogen ration was changed by gas-mixer to keep oxygen partial pressure ($C_o$) from 20% to 100% and a flow velocity of 15 m/s which were controlled by gas supply system (Paker F65). The burning temperatures were measured by thermocouples to reflect changing tendency rather than actual temperature. The examined temperature is ranging from 600 ℃ to 1700 ℃ and the measuring error of 10-thermalcouple made of platinum / platinumrhodium is ±4 ℃. Besides, each type of alloy was tested five times. A digital high-speed video camera (Pco 1200hs, PCO Company, Berlin, Germany) was employed to record the burning characteristics such as burning time, flame height under different burning conditions. The frame intervals and exposure time of the images were 500μs and 10μs, respectively. The resolution of each picture was 768 × 768 pixels. Neutral density filters were placed between the burning samples and the microscope to prevent overwhelming of the light saturation level of the camera. The schematic illustration of the above burning procedure is shown in Fig. S3. A direct current of 5A in a pre-mixed gas was used, and the direct current was turned off immediately after successful ignition. Among that, the main focus of our study is on the burning of the titanium alloy that occurs in air ($Co$ = 20%). Burning test was repeated five times of each kind of alloys to get burning characteristics.

*2.3 Microstructural analyze*

Microstructural investigations were performed to exploit the burned structure by using both optical (Axio Scope-A1 equipped with a Carl Zeiss digital camera) and scanning electron microscopy (SEM, HitachiS-4800) observations. The elemental composition and distribution of burning product were analyzed by using energy dispersive X-ray spectroscopy (EDS, JSM-6700) microanalysis. Reduced area XRD patterns were acquired using a Rigaku Smart Lab (Cu source, operating at 40 kV and 40mA) equipped with a parallel point focus incident beam (CBO-f optic). Certain areas of each sample were selected for analysis. Patterns were collected as 2θ value from 20-90° at a step size of 0.02° at a scan rate of 1.5°per minute. XPS analyses were performed using K-Alpha (Thermo Scientific, the United States). An incident monochromated X-ray beam from the Al target (15 kV, 10 mA) was focused on a 0.7 mm×0.3 mm area of the surface of the sample 45°to the sample surface and the etch time is 20s.

# 3 Results and discussion

## *3.1 Burning characteristics*

Generally, the burning process during a stable burning stage is accompanied by a series of chemical reactions, which are mainly related to the transport and fast reaction of oxygen [20]. During such a process, several burning characteristics are taken into consideration such as burning duration, velocity and flame intensity, which are differentiated by different alloy compositions. Besides, the burning environment is of vital importance for these burning behaviors of alloys [17].

Fig. 2a and Fig. 2b compare the relative burning time and velocity of the two burning resistant alloys under different oxygen partial pressure ($C_O$). It is shown that the burning time and velocity of these two alloys are sensitive to $C_O$, where a high $C_O$ could accelerate burning process and result in long burning time and high burning velocity. These findings agree with those by Shafirovich [21] and Molodetsky [22], where they concluded that burning of titanium alloy is affected by the surface oxygen concentration. It is a reasonable results as a higher $C_O$ will offer higher content of oxygen, which leads to more sufficient burning [23-25] with narrow gap among alloys and thus more burning damage. Comparing with the non-burning resistant TC4 alloy, the burning velocity of Ti40 and Ti14 alloys in air condition ($C_O$ = 20%) are about 53% and 34% slower, respectively, but these differences reduce to 26% and 13% under pure oxygen environment, respectively. This observation suggests that the Ti40 and Ti14 alloys exhibit excellent burn resistance with lower burning velocity and shortest burning time even in high $C_O$ condition, compared with that of TC4 alloy.

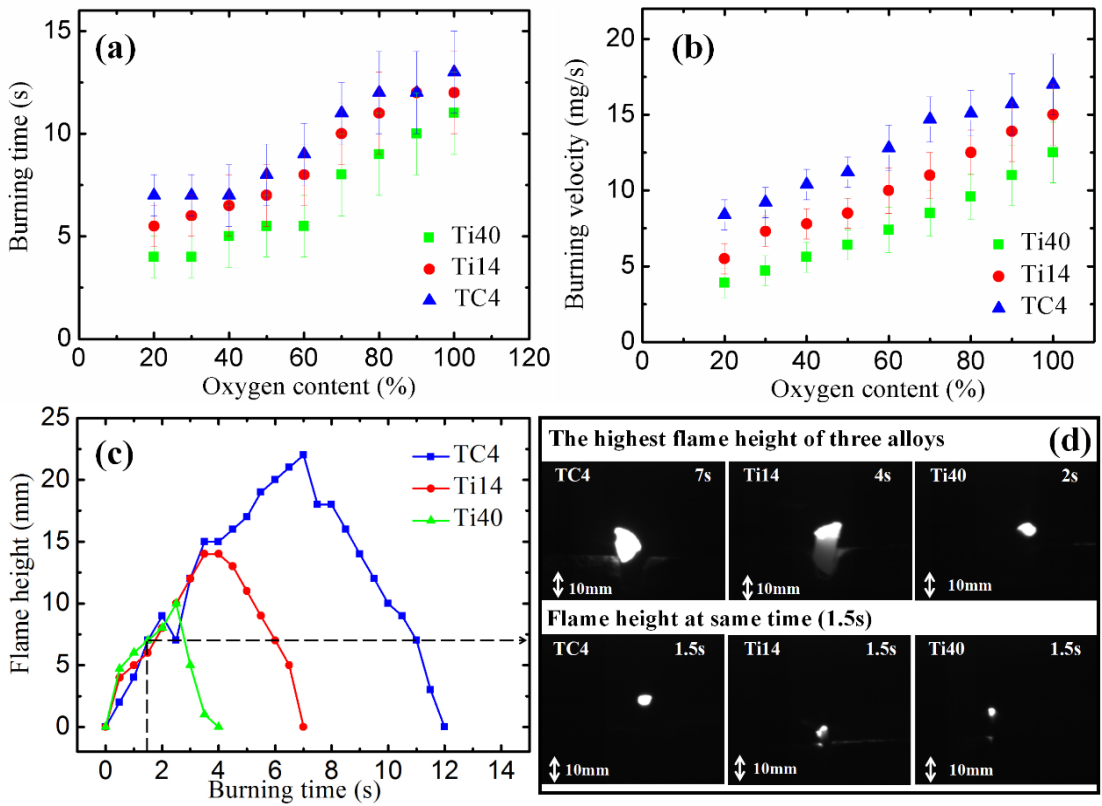

**Fig. 2.** Burning behavior of titanium alloys. (a) The burning duration as a function of the oxygen partial pressure; and (b) the corresponding burning velocity as a function of the oxygen partial pressure; (c) flame height as a function of burning time; and (d) the captured images of Ti alloys at different burning time with an oxygen partial pressure of 20%.

Generally, the reactions taking place in a form of intensive burning produce high brightness, which is directly indicated by the flame height and represents the intensity of burning oxidation reaction [26]. The average flame heights of the alloys with typical burning process during burning process are plotted in Fig. 2c. All samples exhibit a similar parabolic tendency in their flame height with time. By the time when the burning completes, it is found thatTi40 alloy has both the lowest flame height and the shortest total burning duration.Ti14 alloy has a slightly higher flame height than that of Ti40 alloy (about 4mmhigher), which is significantly lower than the maximum flame height measured on TC4 alloy (around 23 mm). According to the captured flame images in Fig. 2d, the highest average flame height of an intensive burning of TC4 alloy occurred at ~1.5s, and little sparks are even observed after 8s. In comparison, Ti40 and Ti14 alloy exhibit similar steady burning from beginning to the end of burning process under the same burning environment. Overall, above results suggest that Ti14 alloy has excellent burning resistant properties compared with that the commercial titanium TC4 titanium alloy, though it is not as good as that of the Ti40 alloy.

### 3.2 Microstructure and composition

To further determine the functionalities of alloying elements Cr, V and Cu, keeping oxygen content at 20% as a normal burning environment to research microstructure and burnt products. Micro-area XRD is conducted at different sample areas to investigate the phase changes induced by burning along the alloy surface and subsurface. The XRD signals are collected from an area of about 2 mm in diameter and thus signify structural characteristics of a local region. The detection is carried out on both alloys at area I and II being close to the burnt surface and on the pristine material, respectively. As illustrated in Fig. 3a and Fig. 4b, the area I shows the clear existence of titanium dioxide phase (rutile $TiO_2$) on top of the burning structure. The formation of stable rutile is due to the high temperature burning environment under the presence of $O_2$. In order to compare the burning product with the substrate matrix, the bottom of the burning structure is selected as the area II, from which a vastly different set of peaks (see Fig. 3a) have been found. For Ti40 alloy, some oxides of Cr and V, e.g., $Cr_2O_3$ phase and $V_2O_5$, exist in the area I but not in the area II. The phase compositions can be extracted for the Ti14 alloy with Cu proportion, which will be discussed further later. Specifically, for Ti14 alloy, the area I shows characteristic peaks of $Cu_2O$ and CuO, and Ti-Cu phase like $Ti_2Cu$ appears as the dominant phase in the burning products in the area I. The results of XRD is corresponding to XPS in Figure 3c and 3f. Burnt surface of Ti14 alloy is mainly composited by O, Ti and Cu, and Ti40 alloy contains a large amount of Ti, O, Cr and V. The certain proportion of different element is shown in Table S1. Carbon C 1s peak at 284.8 eV is used as a reference for charge correction. Moreover, since surface contamination of engine oil produced in sample preparation and treatment, a small amount of carbon is found in burning product zone. In addition, observation of N is caused by contamination of burning atmosphere (mixture of $O_2$ and $N_2$) during burning test but the amount of it is too small to form nitrides detected in XRD. Moreover, the peaks of Cu 2p3/2 and Cu 2p1/2 for the $Cu_xO$ are shown in Fig S4c. Of the two peaks Cu 2p3/2 peak is wider and stronger than Cu 2p1/2. Using the peak shape parameters and peak positions of $Cu^{2+}$ and $Cu^{1+}$ that are obtained from standard maps, the Cu 2p peak was deconvoluted into the $Cu^{2+}$ and $Cu^{1+}$ peaks. Although TiN is observed through XPS except for $TiO_2$, the amount of it is too low to be found through XRD [27, 28].

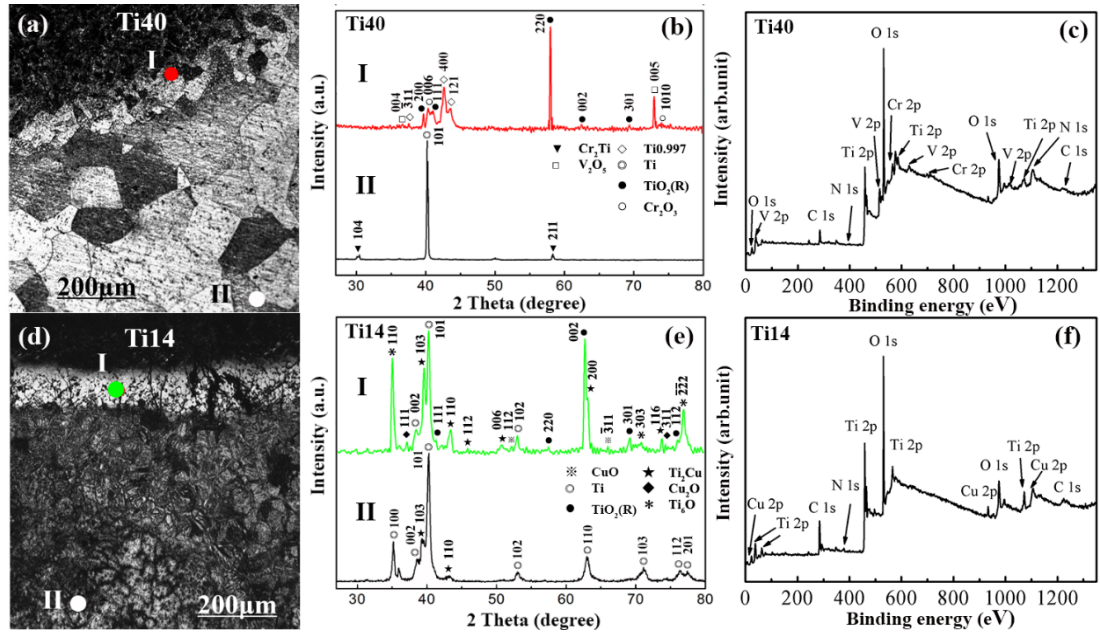

**Fig. 3.** Compositions of the burnt product. (a) Optical microscopic image showing the burned cross-section of Ti40 alloy (under the oxygen partial pressure of $Co = 20\%$); (b) micro-area XRD spectra collected from areas I and II for Ti40 alloy (corresponding to the red and white regions in a); (c) full scan X-ray photoelectron spectra of burnt surface of Ti40 alloy; (d) optical image of the burned cross-section of Ti14 alloy (under the oxygen partial pressure of $Co = 20\%$); (e) the corresponding micro-area XRD data taken from areas I in green and II in white in d; and (f) full scan X-ray photoelectron spectra of burnt surface of Ti14 alloy. Element compositions identified from XPS including $TiO_2$ (PDF No. 04-0551), Ti0.997 (PDF No. 75-0552), Ti (PDF No. 89-3725), $Cr_2Ti$ (PDF No. 49-1716), $Cr_2O_3$ (PDF No. 38-1479), $V_2O_5$ (PDF No. 52-0794), $Cu_2O$ (PDF No. 77-0199), CuO (PDF No. 89-2530), $Ti_2Cu$ (PDF No. 72-0441), $Ti_6O$ (PDF No. 72-1807).

To analyze the elemental distribution and the formation of different product zones, EDS and BSE imaging are carried out. The cross-section morphologies are revealed by SEM and back scattered-electron (BSE) images of the burnt product. Optical microscope image taken across the Ti40 alloy cross section shows a clear division of material phases where the upper section appears to be porous structure $TiO_2$[29] and the lower section presents coarse grains that are associated with the matrix alloy (Fig. 4b). Similar phase zone is also found on post-burning Ti14 alloy. Generally, both Ti40 and Ti14 alloys have similar phase zone configuration, namely burning product zone (BPZ), heat affected zone (HAZ) and the matrix as shown in Fig. 4b and 4f, respectively. For the commercial TC4 alloy, a similar burning structure can also be identified (shown in Supporting Information Fig. S5). Performing EDS line-scanning along the BPZ and HAZ zones in post-burning Ti40 alloy, a variation of element proportion can be observed (Fig. 4a). In BPZ area, both element V and Cr percentages stay relatively constant, whereas the Cr content increases dramatically in the HAZ area. In contrary, the O proportion drops notably from BPZ to HAZ area. This phenomenon signifies a significant oxygen suppression that takes place during burning. According to Fig. 4b, it is believed that a continuous and dense interface is formed between oxide scale and matrix for Ti40 alloy. However, this interface layer has a porous structure (Fig. S5a) with voids and cracks similar as found in TC4 alloy. The porous structure and cracks provide channels for oxidation and accelerate burning processes, which become the major obstacle to obtaining ideal burn resistivity. It is interesting to find that the BPZ area in Ti40 alloy is much larger compared with those in TC4 and Ti14 alloys, which offers a longer distance for oxygen diffusion. The whole BPZ is divided into external surface close to burnt surface and internal surface close to HAZ. Internal surface of BPZ in Ti40 is shown in Fig. 4c with granular substances and exhibits porous microstructure, while compact structure of external surface is shown in Fig 4d. Burnt surface is even, which decreases channels for oxygen diffusion.

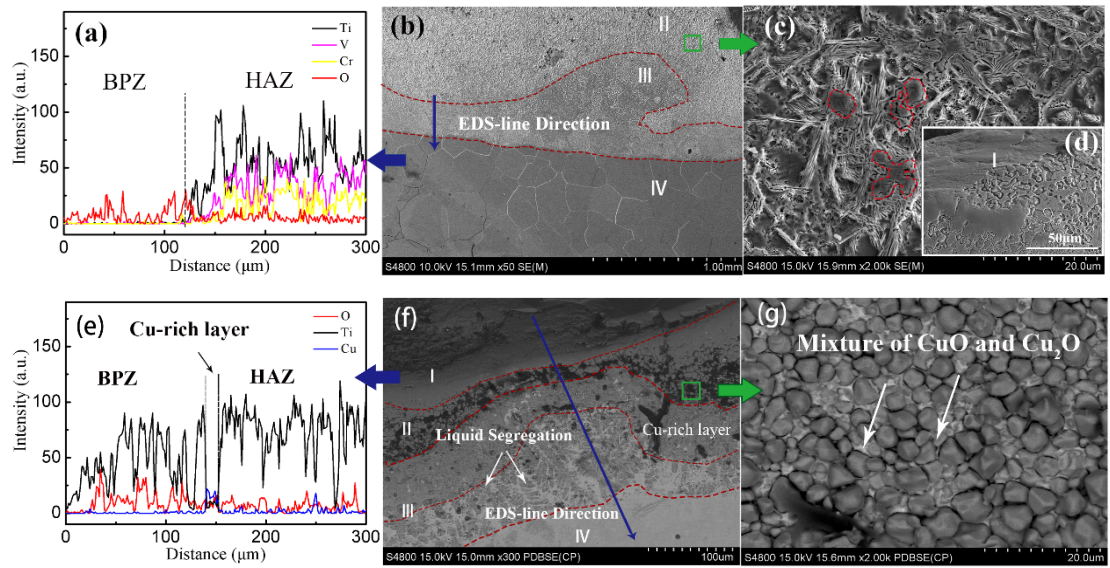

**Fig. 4.** Cross-sectional morphologies of the burnt titanium alloy. The burning product of Ti40 alloy: (a) the change of element distribution along EDS-line in burned structures; (b) SEM image of the burnt structure; (c) magnified image of the burnt product in the burning product zone; (d) SEM image of the burnt surface; The burning product of Ti14 alloy: (e) the change of element distribution along EDS-line in burned structures; (f) SEM image of the burnt structure (with clear segregation phases); (g) a mixture of Cu oxides in the burnt product in the heat affected zone. I, II, III and IV are burned surface, burning product zone (BPZ), heat affected zone (HAZ) and substrate, respectively. The oxygen partial pressure is $Co = 20\%$.

For Ti14 alloy, a mixture phase between CuO and $Cu_2O$ among the $TiO_2$ particles is observed, together with the existence of $Ti_2Cu$ (on the surface of the burned products). As illustrated in Fig. 4f, the Ti14 alloy also exhibit a porous microstructure with plenty of voids due to the presence of $TiO_2$ (which intrinsically possess a loosened and porous structure [30,31] ) after burning. These voids act as transporting channels for oxygen during burning. The existences of copper oxides will increase the density of the surface structure/layer, and thus hinders the oxygen diffusion. It is worth to note that segregation of liquid Cu form HAZ to BPZ occurs during burning, which forms a Cu-rich layer between BPZ and HAZ areas. A clear detection of Cu content change is derived by running EDS line scanning along the burned surface shown in Fig. 4e. A boost in Cu proportion is seen along the cross section as highlighted by black arrow (Fig. 4g). It is also found that the content of O in HAZ area is lower than that in BPZ area. Compared with other alloys, all burning affected areas of TC4 alloy are small with coarse grains with clearly found cracks, and oxygen appears the most abundant in HAZ area of TC4 alloy. Overall, above observations show that Cu is preventing the oxidized diffusion and restrains further burning in Ti14 alloy.

*3.3 Functionalities of alloying element Cr/V and Cu*

According to above discussions, the burning of titanium alloy is actually controlled by oxidizing reaction and oxygen diffusion [32]. At the initial stage, oxidizing reaction is the main force to drive burning on the surface which forms various burning products [33]. As burning continues, oxygen gradually diffuses into the matrix through oxide layer to maintain burning process. The different burn resistant capacity of Ti14 and Ti40 alloy arises from the alloying element composition. Based on the burning characteristics and burned microstructures, the burning resistant mechanisms induced by the alloying element Cr/V and Cu can be explained by the following models.

For both Ti40 and Ti14 alloy, a thin and porous Ti oxides layer is formed during the initial stage of burning, and the growth of the oxide proceeds by the outward diffusion of metal cations and the inward diffusion of oxygen. The diffusion rate of the ions is determined by the size of cation. As depicted in Fig. 5, the small cation is easier to diffuse in the oxide scale than the large cation. The ionic radius of Ti, V and Cr are 132 pm, 122 pm and 118 pm, respectively, which leads to the following diffusion rate order Ti< V< Cr [22]. As the burning oxidation progresses, the outward diffusion of Cr cations (together with the dissolution of $TiO_2$), increases the Cr concentration at the oxide/metal interface, leading to the distribution of Cr mainly in HAZ area. The outward diffusion of V and Cr leads to the formation of $Cr_2O_3/V_2O_5$ composite among the porous oxide layer, which reinforces the oxide scale with a more compact structure. Since the density of $Cr_2O_3$ (4.89g/cm$^3$) is higher than that of $TiO_2$ (3.62g/cm$^3$), the Ti cations in the matrix are difficult to transfer outwards through $Cr_2O_3$ layer and fail to supply titanium source at the surface for continuous burning. This compact structure is beneficial for the isolation of the oxygen transport and results in low burning velocity (as shown in Fig. 2). Further evidence is affirmed from EDS mapping in Fig. 6b, from which the content of O in HAZ area is the lowest compared with Ti14 and TC4 alloy. Besides, evaporation of $V_2O_5$ (melting point is 690℃ shown in Table S2) can take away a lot of heat that is supplied to the burning reaction [34].

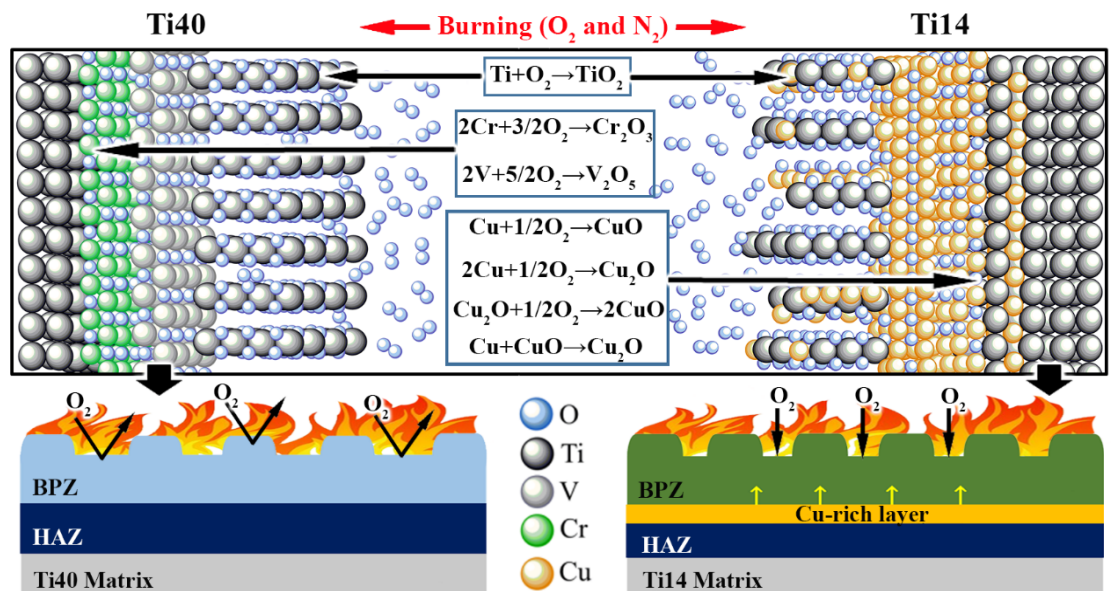

**Fig. 5.** Schematic model of the structure of Ti40 and Ti14 during burning.

The element distribution of Ti14 alloy is relatively homogenous compared with Ti40 alloy. During oxidation process, the Cu elements segregate to the interface of the burning product and matrix of Ti14 alloy, resulting in a Cu-rich layer [35,36] as seen from the EDS-mapping in Fig. 6a. Evaporation of CuO and $Cu_2O$ (melting point is 1065 °C and 1235 °C, respectively, see Table S2) under high temperature results in the disappearance of Cu at the surface region. Being different from the blocking effect of the mixed $V_2O_5$ and $Cr_2O_3$ oxide layer in Ti40 alloy, a continuous oxidizing reaction (see Fig. 5) between Cu and O occurs in the Cu-rich interface layer. Unlike the diffusion ability of Cr and V, the radius of Cu is 145 pm, which can hardly block Ti transfer towards the surface. After burning, Ti14 alloy is still consisted of $α$-Ti and $Ti_2Cu$, there is no phase transformation from $α$ to $β$ phase of Ti during cooling process. The main function of Cu on burn resistance is thus achieved through the reaction with oxygen diffused form BPZ through a Cu-rich layer. Moreover, Cu-rich layer is form as Cu-rich phase increase distance between oxygen and titanium to influence oxygen diffusion. Thus, a low concentration of oxygen in HAZ area is observed with the existence of Cu in BPZ area (as shown in Fig. 4a).

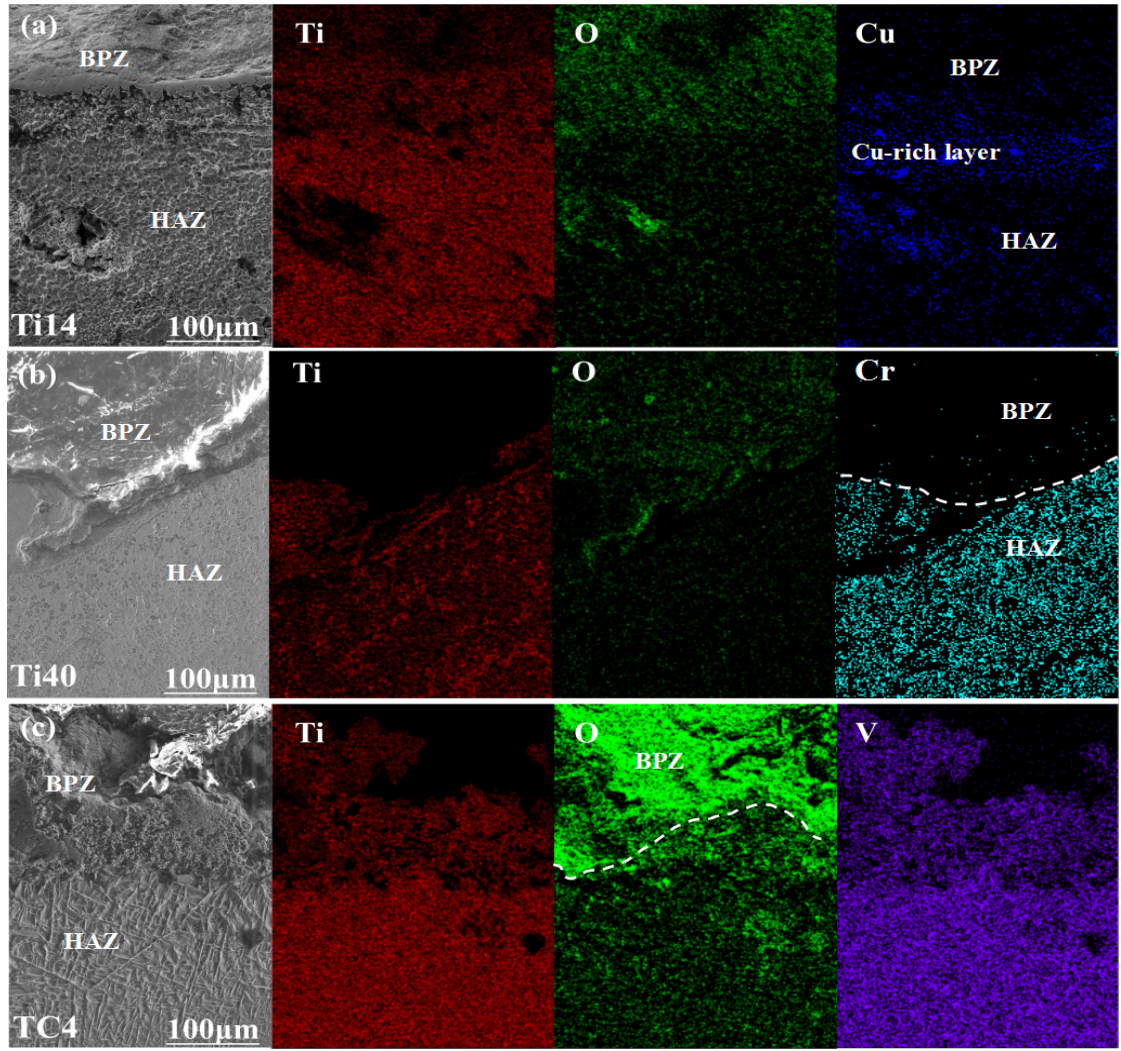

**Fig. 6.** SEM (BSE) images and element maps of titanium alloys after burning. (a) Ti14 alloy; (b) Ti40 alloy; and (c) TC4 alloy.

In particular, the formation of Cu-rich layer is the result from segregation of Ti-Cu alloy caused by temperature gradient. It is known that the $Ti_2Cu$ precipitates in Ti14 alloy change to liquid form, which makes the alloy next to the burning surface consists of both liquid and solid phases, i.e., a semi-solid region (given the melting point of $Ti_2Cu$ is 990ºC [16]). In the burning processes, more liquid segregates from matrix to semi-solid region to form Cu-rich region (Fig. 4f and Fig. 4g), instead of keeping the titanium alloy reacts with oxygen continuously. Distance between Ti and O is increased by Cu-rich layer and more energy is required to diffusion through Cu-rich layer, therefore, diffusion processes of Ti and O are blocked. The segregation of Ti-Cu alloy introduces a liquid lubricating effect to Ti14 alloy as is seen in semi-solid forging process [37,38]. It is expected that the molten liquid phase can act as lubrication on the surface to reduce the burning possibility as caused by friction, which is the unique characteristic of Ti14 burn resistant alloy and requests further investigation.

In summary, it is shown that both Ti40 and Ti14 have similar burn resistant functionalities, while the resistant mechanisms are totally different. The major role of V and Cr in Ti40 alloy is to form $V_2O_5$ and $Cr_2O_3$ oxides, which enhance the density of the oxide layer that impedes the diffusion of oxygen. On the contrary, a Cu-rich layer is formed in Ti14 alloy by the diffusion of Cu, and Cu "sacrifices" itself to react with $O_2$. Based on the burning characteristics and burning structure, although Cu could provide burn resistance, it is not as effective as that of V or Cr. However, in some undemanding applications for burn resistance (such as engine brackets, auto engines, and replaceable materials on friction-prone surfaces), it can offer a cost-effective way to design burn resistant titanium alloys.

## 4. Conclusions

This work thoroughly studied the burning resistant mechanisms for two representative burning resistant alloys, Ti40 and Ti14 alloys keeping in same burning condition through direct current simulation burning method. Compared with the non-burn resistant TC4 (Ti-6Al-4V) alloy, both Ti40 and Ti14 alloys possess better burn resistant property. Specifically, Ti40 alloy exhibits a better burn resistant capability with lower burning velocity and flame height. It is found that V and Cr in Ti40 alloy forms $V_2O_5$ and $Cr_2O_3$ oxides, which enhance the density of the oxide layer that impedes the diffusion of oxygen. On the contrary, a Cu-rich layer is formed in Ti14 alloy by the diffusion of Cu to reduce exposure of Ti to oxygen and more energy is required for oxygen to get into substrate. Meanwhile, a portion of Ti were replaced by Cu to react with oxygen to form CuO and $Cu_2O$, which consumes a certain amount of oxygen. Additionally, the liquid phase of $Ti_2Cu$ in Ti14 alloy during dramatic impact will cause lubricating effect to improve the burn resistance. This work suggests that replacing V or Cr by Cu, the titanium alloy could still attain certain burn resistant function, which could greatly reduce the high cost for certain undemanding applications like materials of auto-engines and brackets for vehicles and replaceable materials on friction-prone surface in chemistry engineering. The established understanding offers practical guidance for the design and application of cost-effective burning resistant titanium alloy.

**Acknowledgments**

The authors acknowledge financial supported by National Key Research and Development Program of China (No.2016YFB0700301, No.2016YFB1100103), National Natural Science Foundation of China (grant no. 51471136), Postdoctoral Science Foundation of ShaanXi province (2017BSHYDZZ01), and the Special Fund for Basic Scientific Research of Central Colleges (300102318205, 300102318206 and 300102318209). The Y. Chen, A. Bo, H. Zhan and Y. Gu authors acknowledge Australian Research Council (DP170102861) and technical supports from Central Analytical Research Facility (CARF) of Queensland University of Technology (QUT).

## Author Contributions

Y. Chen and W. Yang contribute equally to this work. Y. Chen, W. Yang, Q. Zhao, M. Wan and F. Zhang planned and conducted all experiments. Y. Chen, W. Yang, H. Zhan, A. Bo and Y. Zhao examined the experimental results, performed data analyses and wrote the article draft together. Y. Chen, F. Zhang, A. Bo and Y. Gu were involved in the data analysis, discussion and writing.

## Conflicts of Interest

The authors declare no conflict of interest.